\documentclass{CHEP2006}
\usepackage{graphicx}

\begin{document}
\title{COMPUTATION OF NEARLY EXACT 3D ELECTROSTATIC FIELD IN GAS IONIZATION
DETECTORS}

\author{N. Majumdar, S. Mukhopadhyay\\ Saha Institute of Nuclear Physics\\
1/AF, Bidhannagar, Kolkata-700064, India\thanks{nayana.majumdar@saha.ac.in}}

\maketitle

\begin{abstract}
The three-dimensional electrostatic field configuration in gas ionization
detectors has been simulated using an efficient and precise nearly exact
boundary element
method (NEBEM) solver set up to solve an integral equation of the first kind.
This recently proposed formulation of BEM and the resulting
solver use exact analytic integral of Green function to compute the
electrostatic potential for charge distribution on flat surfaces satisfying
Poisson's
equation. As a result, extremely precise results have been obtained despite
the use of relatively coarse discretization leading to successful validation
against analytical results
available for two-dimensional MWPCs. Significant three dimensional effects
have been observed in the electrostatic field configuration and also on the
force experienced by the anode wires of MWPCs.
In order to illustrate the applicability of the
NEBEM solver for detector geometries having multiple dielectrics and degenerate
surfaces, it has been validated against
available FEM and BEM numerical solutions for similar geometries.
\end{abstract}

\section{INTRODUCTION}

Electrostatic forces play a major role in determining gas detector performance.
Hence, a thorough understanding of electrostatic properties of gas detectors is
of critical importance in the design and operation phases. Computation of
electrostatic field is conventionally carried out using one of the following
options: analytical approach, finite-element method (FEM), finite-difference
approach and boundary element method (BEM). While the first of these
possibilities offers the most precise estimation, it is known to be severely
restricted in handling of complicated and realistic detector configurations. The
FEM is the most popular being capable of producing reasonable results for
almost arbitrary geometries \cite{Garfield}. However, for the present task, the FEM is not
a suitable candidate because through this method it is possible to calculate
potentials only at specific nodal points in the detector volume. For non-nodal
points, interpolation becomes a necessity reducing the accuracy significantly.
Moreover, numerical differentiation for obtaining field gradients leads to
unacceptable electric field values in regions where the gradients change
rapidly, for example, near the anode wires of an MWPC. The BEM, despite being
capable of yielding nominally exact results while working on a reduced
dimensional space(unknowns only on surfaces rather than volumes), suffers from
inaccuracies near the boundaries. They also necessitate relatively complicated
mathematics.

In this work, we propose the application of a recently formulated Nearly Exact
BEM solver (NEBEM) \cite{Mukhopadhyay05} for the computation of electrostatic
field in gas detectors. The BEM is a numerical implementation of boundary
integral equation (BIE) based on Green's function method by discretization of
the boundary only. For electrostatic problems, the BIE can be expressed as
\begin{equation}
\phi(\vec{r}) = \int_S{G(\vec{r}, {\vec{r}}^\prime) \rho({\vec{r}}^\prime)
			dS^\prime}
\end{equation}
where $\phi(\vec{r})$ represents the known voltage at $\vec{r}$, $\rho(\vec{r}^\prime)$ represents the charge density at ${\vec{r}}^\prime$,
$G(\vec{r}, \vec{r}^\prime) = 1 / 4 \pi \epsilon | \vec{r} - \vec{r}^\prime |$
and $\epsilon$ is the permittivity of the medium. In order to develop a solver
based on this approach, the charge carrying boundaries are segmented on which
unknown uniform charge densities ($\rho$) are assumed to be distributed. The
unknown $\rho$ and the known $\phi$ are related through the influence matrix
$\matrix{A}$
\begin{equation}
\matrix{A} . \rho = \phi
\label{eq:Matrix}
\end{equation}
where $A_{ij}$ of $\matrix{A}$ represents the potential at the mid point of
segment $i$ due to a unit charge density distribution at the segment $j$. By
solving (\ref{eq:Matrix}), the unknown $\rho$ can be obtained.

Major approximations are made while computing the influences of the
singularities which are modeled by a sum of known basis functions with constant
unknown coefficients. These approximations ultimately lead to the infamous
numerical boundary layer due to which the solution at close proximity of the
boundary elements is severely inaccurate.

\section{PRESENT APPROACH}
In the present approach, we have proposed to use the analytic expressions of
potential and force field due to a uniform distribution of singularity on a
flat rectangular surface in order to compute highly accurate influence
coefficients for the $\matrix{A}$ and for calculations subsequent to the
solution of $\rho$. By adopting such an approach, it is possible to improve
up on the above-mentioned assumption of singularities concentrated at nodal
points and move to uniform charge density distributed on elements. In general,
the potential $\phi$ at a point $(X,Y,Z)$ in free space due to uniform source
distributed on a rectangular flat surface on the $XZ$ plane and having corners
at $(x_1, 0, z_1)$ and $(x_2, 0, z_2)$ is known to be a multiple of
\begin{equation}
\phi(X,Y,Z) = \int_{z_1}^{z_2} \int_{x_1}^{x_2}
\frac{dx dz}{\sqrt{ (X-x)^2 + Y^2 + (Z-z)^2}}
\end{equation}
Closed form expressions for the above integration and also for the force
vectors have been obtained and used as the foundation expressions of the NEBEM
solver described in \cite{Mukhopadhyay05}. It should be noted that these new
foundation expressions are valid throughout the physical domain including the
close proximity of the singular surfaces. In the present work, we have extended
the NEBEM solver to solve the electrostatic problem of gas detectors as follows.

For gas detectors, it is also be useful to model the electrostatics of a wire.
For this purpose, we have considered the wires to be thin in comparison to the
distance at which the electrostatic properties are being estimated. Under this
assumption, the potential $\phi$ at a point $(X,Y,Z)$ due to a wire element
along the $Z$ axis of radius $r$ and length $L$ is as follows:
\begin{equation}
\phi = 2 \pi r
log(\frac{\sqrt{X^2+Y^2+(Z-h)^2} - (Z-h)}{\sqrt{X^2+Y^2+(Z-h)^2} - (Z+h)})
\label{eq:WirePot}
\end{equation}
where $r$ is the radius of the wire and $h$ is the half-length of the wire
element.
Similar expressions for the force field components have also been obtained which
are not provided here due to the lack of space. In addition to the expressions
given in \cite{Mukhopadhyay05}, (\ref{eq:WirePot}) and the companion force field
expressions have been incorporated in the NEBEM solver to compute the electric
potential and field in ionization detectors.

\section{NUMERICAL IMPLEMENTATION}
In this work, we are going to present results related to Iarocci tube, MWPC
and RPC. These devices have flat surfaces as their boundaries. The Iarocci
tube and the MWPC have anode wires, in addition to the flat conducting surfaces.
We have considered Iarocci tubes and MWPCs of various cross-sections ($5mm\times
5mm$ to $16mm\times 16mm$), lengths ($5mm$ to $160mm$) and anode wire diameters
($20 \mu m$ to $100 \mu m$). The anode wires have been assumed to be held at
$1000V$ whereas the other surfaces are assumed to be grounded.
The flat surfaces in all these detectors have been
segmented in to $21 \times 21$ elements. The anode wires have been modeled as
cylindrical polygons of 12 sides. Along the axial direction, these cylinders
have been segmented in to 21 sub-divisions. These are once again flat surfaces
and,
thus, have been modeled by the same approach as discussed above. The anode
wires
have also been represented as thin wires for which (\ref{eq:WirePot}) and
related expressions have performed as foundation expressions. The maximum number
of elements with a wire representation for an Iarocci tube has been 1785 while
that for a MWPC with five wires have been 1869. Under the assumption of
cylindrical polygons, these have increased to 2016 and 3024. These numbers are
quite modest and the resulting system can be easily solved on modest desktop
computer. Please note that the solver solves a complete three-dimensional
problem. We have executed our codes on a Pentium IV machine with 2GB RAM running
Fedora Core 3 and it took approximately half an hour of user time to solve the
most complicated problem.

\section{RESULTS}
\begin{figure}[htb]
\centering
\includegraphics*[width=65mm]{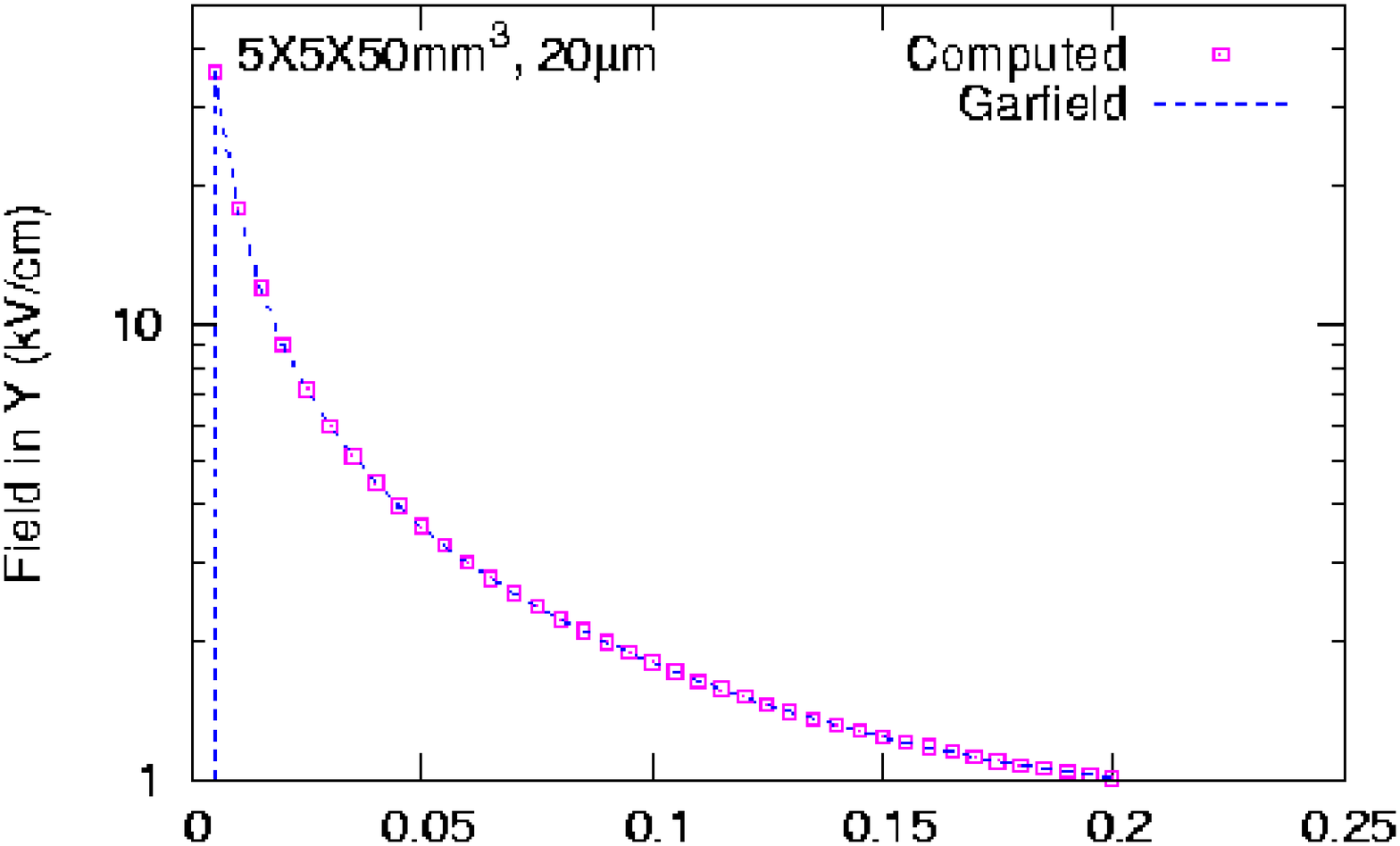}
\caption{Comparison of electric field at the midplane}
\label{fig:EFIarocci1}
\end{figure}

\begin{figure}[htb]
\centering
\includegraphics*[width=65mm]{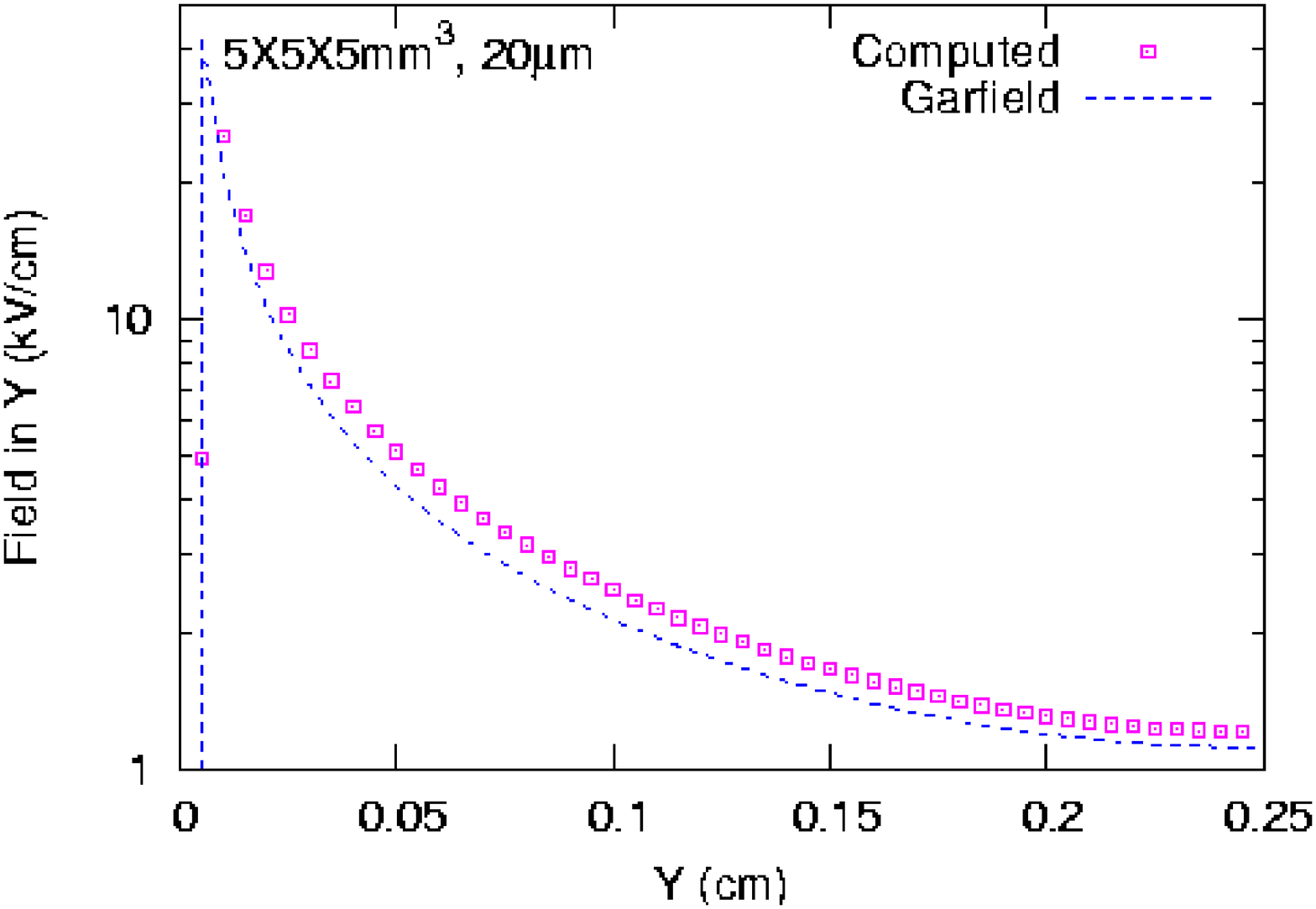}
\caption{Comparison of electric field at the midplane}
\label{fig:EFIarocci2}
\end{figure}

\begin{figure}[htb]
\centering
\includegraphics*[width=65mm]{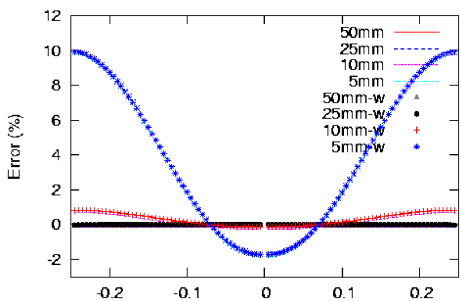}
\caption{Deviation of electric field at the midplane}
\label{fig:Error1}
\end{figure}

\begin{figure}[htb]
\centering
\includegraphics*[width=65mm]{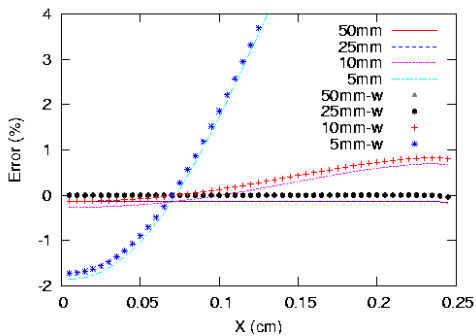}
\caption{Close up of deviation of electric field at midplane}
\label{fig:Error2}
\end{figure}

In this section, we will concentrate mostly on the electric field, rather than
the potential, to save space. In Fig.\ref{fig:EFIarocci1}, we have compared the
midplane electric fields computed by the present approach and analytic electric
field computed by \cite{Garfield} for a $5mm\times5mm\times50mm$ Iarocci
tube. It can be seen that the two results match perfectly well. However, when
the length of the tube is reduced to $5mm$, the three-dimensionality of the
device becomes relevant even at the mid-plane. Thus, in Fig.\ref{fig:EFIarocci2}
the difference between the two values have become apparent.
In order to
illustrate the effect of three-dimensionality, we have presented the relative
deviation defined as $Error(\%)=\frac{Garfield-Present}{Garfield}\times100$ for
various Iarocci detectors in Figs.\ref{fig:Error1} and \ref{fig:Error2}. In
these figures, results for wires represented as thin wires, as well as
cylindrical polygons have also been presented. All the electric fields are at
the mid-plane of the 3D detector, the length of which has been varied from 50mm
to 5mm. It can be seen that the deviation becomes apparent (close to 1\%) as the
detector length becomes twice the side of the square detector. When the length
is the same as that of the side, the error is as large as 10\% near the cathode
surfaces. In the latter case, it is more than 2\% near the anode wire, which is
huge considering the magnitude of the electric field near the anode wire. Fig.
\ref{fig:Error2} illustrates the same point more clearly and, in addition,
shows the difference between the two representations of wire. It may be noted
that electric field computed by the wire representation seems to be consistently
higher than that obtained using the polygon representation.

\begin{figure}[htb]
\centering
\includegraphics*[width=65mm]{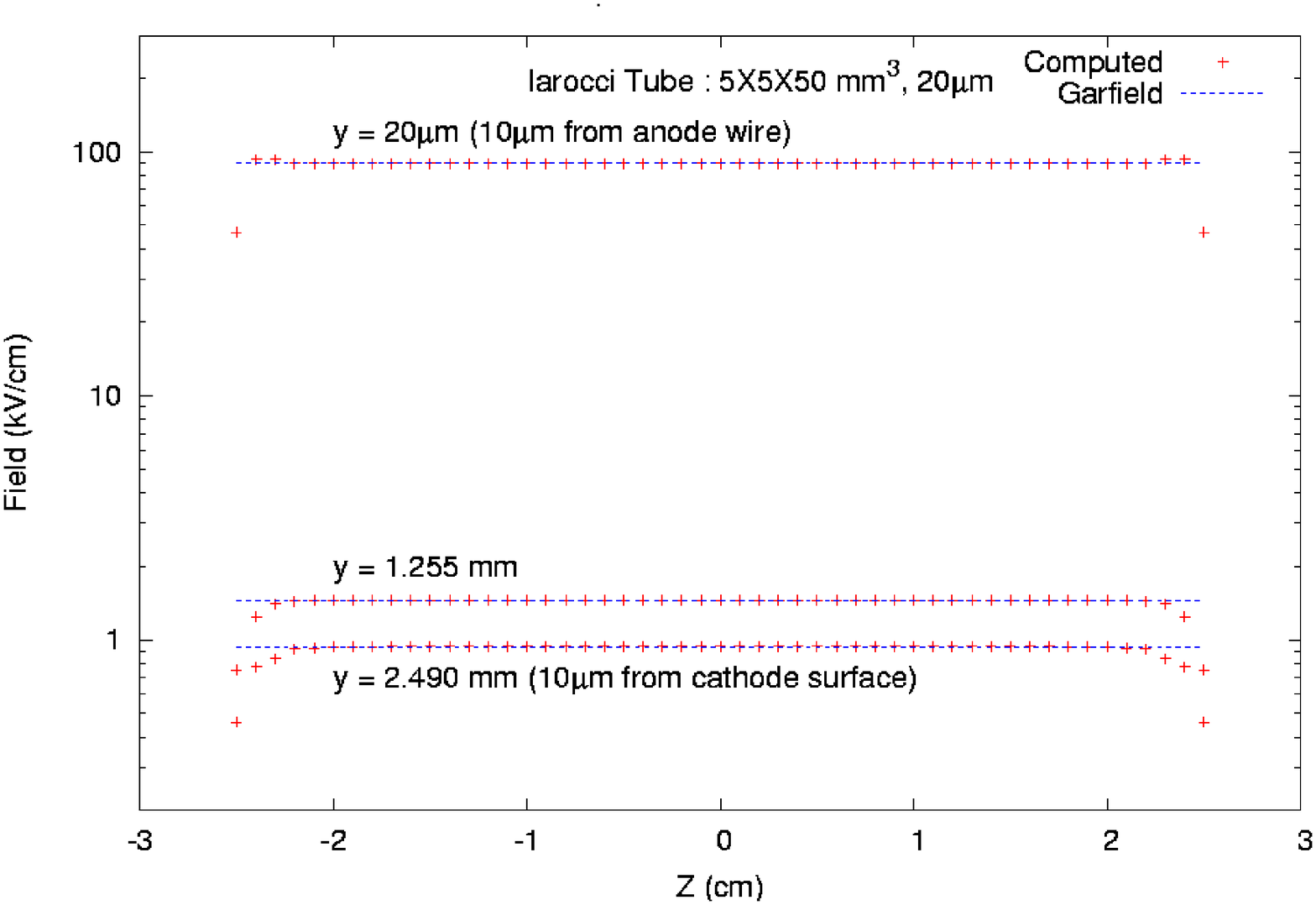}
\caption{Comparison of normal field along the axial direction}
\label{fig:Eyaxial}
\end{figure}

Next, we present the normal electric field computed along the axial direction
of the Iarocci tube in Fig.\ref{fig:Eyaxial}. The reference straight line has
been provided by using \cite{Garfield} and is the analytic solution for a 2D
tube. It can be seen that the 2D value remains valid for almost 85\% of the
detector along the axial direction. However, in the remaining 15\%, 3D effects
are quite prominent and hence, non-negligible. There is one more very important
point to be noted. Because of the nature of FEM, it almost produces oscillatory
results in potential and field (less in the former and more in the latter) near
the cathode surface and anode wires. But these are the locations where the
results need to be most accurate! By using NEBEM, it has been possible to
produce perfectly smooth results without any hint of oscillation. This precision
can only be attributed to the exact nature of the foundation expressions of this
solver. This remarkable feature of the present solutions should allow more
realistic estimation of detector behavior in any situation.

\begin{figure}[htb]
\centering
\includegraphics*[width=65mm]{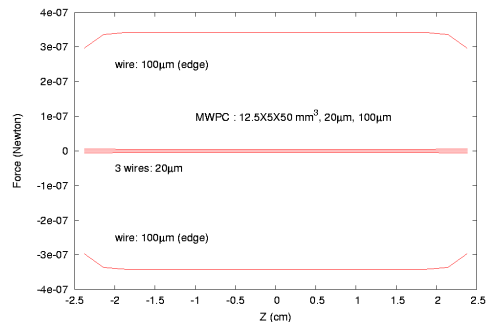}
\caption{Variation of horizontal force along the axial direction}
\label{fig:Force}
\end{figure}

Similar validation and comparisons have been carried out for an MWPC having
five anode wires with a wire pitch of $2.5mm$ leading to similar
conclusions. There is one feature which is considered to be important mainly for
MWPCs, namely, the electrostatic force that acts on the anode wires in an MWPC,
especially, its positional variation. In Fig.\ref{fig:Force},
we present the variation of the horizontal force acting on the different anode
wires in a five-wire MWPC as we move along the length of the wire. It should be
mentioned here that the edge wires in the presented case are of $100 \mu m$
diameter, while those inside are of $20 \mu m$.
It is seen that the horizontal force becomes quite considerable as we move
towards the edge of the detector despite the use of larger diameter wire as
guard wires.

In the following Fig.\ref{fig:EFcontrMWPC}, we have presented the normal
electric field contour at the mid-plane of the detector. The sharp increase in
the magnitude of this field is quite evident near the wire locations.
\begin{figure}[htb]
\centering
\includegraphics*[width=65mm]{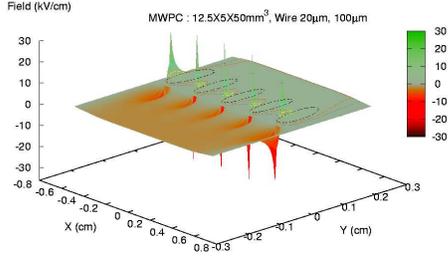}
\caption{Surface and contour plots of the normal electric field}
\label{fig:EFcontrMWPC}
\end{figure}

Several detector configurations in current use such as MSGC, RPC, GEM have
multiple dielectric configuration. They also have extremely thin charged
surfaces such as the graphite coatings of RPCs. Such surfaces may necessitate
treatment of degenerate surfaces. As a result, in the following, we have tried
to validate the NEBEM solver by computing the electrostatic properties of
such configurations and comparing them with FEM and Dual BEM (DBEM) results
given in \cite{Chyuan} where the problem geometry is discussed in detail. In the
following Tables \ref{table:t1}, \ref{table:t2} and \ref{table:t3}, we present
the comparison among the potentials as computed by the above three approaches.
It is obvious from these results that the
agreement is excellent despite the use of only $11\times11$ discretization of
the surfaces. Please note that in the tables $R$ denotes the ratio between the
permittivity of two slabs placed on top of another (upper/lower) and the 
locations are expressed in microns.
\begin{table}[hbt]
\begin{center}
\caption{Comparison for R = 10}
\begin{tabular}{|l|c|c|c|}
\hline \textbf{Location} & \textbf{FEM} & \textbf{DBEM} & \textbf{NEBEM}\\ \hline
18.0,3.0	& 0.1723103	& 0.17302	& 0.1740844 \\
4.0,9.0		& 0.2809692	& 0.27448	& 0.2807477 \\
25.0,16.0	& 0.6000305	& 0.59607	& 0.5991884 \\
5.0,17.0	& 0.679071	& 0.67492	& 0.6785017 \\ \hline
\end{tabular}
\label{table:t1}
\end{center}
\end{table}

\begin{table}[hbt]
\begin{center}
\caption{Comparison for R = 0.1}
\begin{tabular}{|l|c|c|c|}
\hline \textbf{Location} & \textbf{FEM} & \textbf{DBEM} & \textbf{NEBEM}\\ \hline
18.0,3.0	& 0.01741943	& 0.017302	& 0.0171752 \\
4.0,9.0		& 0.0281006	& 0.027448	& 0.0286358 \\
25.0,16.0	& 0.4883313	& 0.480640	& 0.4828946 \\
5.0,17.0	& 0.5929200	& 0.589690	& 0.5926387 \\ \hline
\end{tabular}
\label{table:t2}
\end{center}
\end{table}

\begin{table}[hbt]
\begin{center}
\caption{Comparison for R = 10}
\begin{tabular}{|l|c|c|c|}
\hline \textbf{Location} & \textbf{FEM} & \textbf{DBEM} & \textbf{NEBEM}\\ \hline
24.0,16.5	& 0.514489	& 0.52181	& 0.5247903 \\
6.5,12.0	& 0.230147	& 0.23801	& 0.2398346 \\
22.5,6.0	& 0.3638855	& 0.34638	& 0.3451232 \\
4.0,3.5		& 0.1108643	& 0.10623	& 0.1058357 \\ \hline
\end{tabular}
\label{table:t3}
\end{center}
\end{table}

\begin{figure}[htb]
\centering
\includegraphics*[width=65mm]{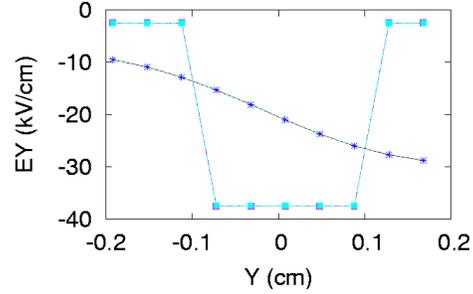}
\caption{Electric field in an RPC}
\label{fig:rpc}
\end{figure}

Finally, in Fig.\ref{fig:rpc}, we have presented the electric field
configuration of an RPC computed using NEBEM. The RPC is assumed to have 2mm gas
gap enclosed by two 2mm glass. On the two outer surfaces of the glass, graphite
coating has been applied. The upper graphite coating has been raised to 8kV,
while the lower one is grounded. The RPC is 10mm wide, enclosed on the left and
right sides by glass and has a similar spacer in the middle. The variation of
the electric field along the vertical direction (Y) on the midplane at various
X locations have been presented. It can be seen that the normal electric field
rises up to around 40kV/cm in the gas gap of the RPC.

\section{CONCLUSION}
Thus, it may be concluded that
\begin{Itemize}
\item
{Precise computation of three-dimensional surface charge density, potential and
electrostatic field has been carried out for gas ionization detectors using the
Nearly Exact BEM (NEBEM) solver.}
\item
{The NEBEM yields precise results for a very wide range of realistic
electrostatic configurations including multiple dielectric systems.}
\end{Itemize}

As future plans we have the following aspects
\begin{Itemize}
\item
{Optimization of the NEBEM solver.}
\item
{Multiphysics nature of NEBEM needs to be explored. After all, the foundation
expressions represent new solutions to the Poisson equation which is one of the
most important equations in physics!}
\item{Applications to other areas will be explored}
\end{Itemize}

\section{ACKNOWLEDGEMENTS}
We would like to acknowledge the help and encouragement of Prof. Bikash Sinha,
Director, SINP and Prof. Sudeb Bhattacharya, Head, Nuclear and Atomic Physics,
SINP throughout the period of this work.

\end{document}